\documentclass[10pt,aps,prd,letter,floatfix,showpacs,twocolumn,superscriptaddress]{revtex4-2}

\usepackage{graphicx}
\usepackage{amsmath, amsfonts, amssymb, bm}

\usepackage{amstext}
\usepackage{mathrsfs}
\usepackage{hyperref}
\usepackage{color}

\usepackage{pdfpages}
\usepackage{pgffor}
\usepackage{etoolbox} 

\makeatletter
\patchcmd{\@outputpage@head}{\@ifx{\LS@rot\@undefined}{}{\LS@rot}}{}{}{}
\makeatother

\begin{document}

\title{Enhanced quantum radiation with flying-focus laser pulses}
\author{Martin S. Formanek}
\email{martin.formanek@eli-beams.eu}
\affiliation{ELI Beamlines Facility, The Extreme Light Infrastructure ERIC, 252 41 Doln\'{i} B\v{r}e\v{z}any, Czech Republic}
\author{John P. Palastro}
\affiliation{Laboratory for Laser Energetics, University of Rochester, Rochester, New York 14623, USA}
\author{Dillon Ramsey}
\affiliation{Laboratory for Laser Energetics, University of Rochester, Rochester, New York 14623, USA}
\author{Antonino Di Piazza}
\affiliation{Department of Physics and Astronomy, University of Rochester, Rochester, New York 14627, USA}
\affiliation{Laboratory for Laser Energetics, University of Rochester, Rochester, New York 14623, USA}

\begin{abstract}
The emission of a photon by an electron in an intense laser field is one of the most fundamental processes in electrodynamics and underlies the many applications that utilize high-energy photon beams. This process is typically studied for electrons colliding head-on with a stationary-focus laser pulse. Here, we show that the energy lost by electrons in the quantum regime and the yield of emitted photons can be substantially increased by replacing a stationary-focus pulse with an equal-energy flying-focus pulse whose focus co-propagates with the electrons. These advantages of the flying focus result from the energy loss and the photon yield  scaling more favorably with the interaction time than the laser intensity in the quantum regime, with the latter also holding in the classical regime. Monte Carlo simulations of electrons colliding with equal-energy stationary and flying-focus laser pulses demonstrate these advantages. 

\end{abstract}

\maketitle

\section{Introduction}
Modern high-power lasers enable direct probing of strong-field quantum electrodynamics with optical light \cite{E144:1996enr, Burke:1997ew, Cole:2017zca, Poder:2017dpw, mirzaie2024, los2024observation}. Strong-field quantum electrodynamics (SFQED) refers to processes that occur in electromagnetic fields with amplitudes of the order of the Schwinger critical field $F_\text{cr} = m^2/|e| = 1.3 \times 10^{16}$ V/cm = $4.4\times 10^{13}$ G \cite{DiPiazza:2011tq,Burton:2014wsa,Gonoskov:2021hwf,Fedotov:2022ely}. Here, $e < 0$ is the electron charge, $m$ its mass, and units $\hbar = c =\epsilon_0= 1$ are employed throughout. The corresponding intensity is $I_\text{cr} = 4.6 \times 10^{29}$ W/cm$^2$. To date, the highest intensity achieved by a high-power laser in the laboratory frame is about six orders of magnitude below this value \cite{Yoon:2021ony}. Nevertheless, the peak intensity experienced by an ultrarelativistic electron in its rest frame can readily exceed $I_\text{cr}$. The key parameter that determines the strength of a SFQED process is $\chi = \sqrt{-(F^{\mu\nu}p_\nu)^2}/mF_\text{cr}$, where $p_\nu$ is the electron four-momentum, $F^{\mu\nu}$ is the electromagnetic tensor of the laser field, and the metric is $\eta^{\mu\nu} = \text{diag}(+1,-1,-1,-1)$. Because $\chi$ is a Lorentz-invariant quantity, it is directly proportional to the electric field experienced by an electron in its rest frame. As a result, a value of $\chi \gtrsim 1$ can be achieved experimentally by colliding available beams of multi-GeV electrons \cite{Gonsalves:2019wnc,Abramowicz_2019,Meuren_2020,Picksley:2024cdd} with existing high-power laser pulses.

Space-time structured laser pulses, such as the flying focus (FF), offer a new paradigm for probing SFQED processes. FF pulses feature a programmable velocity focus that moves independently of the group velocity \cite{Sainte-Marie_2017,Froula_2018,Turnbull_2018}. The moving intensity peak formed by the FF can travel distances much longer than a Rayleigh range while maintaining a near-constant spatiotemporal profile \cite{Froula_2018,Palastro_2018,Turnbull_2018,Howard_2019,simpson2020nonlinear,Ramsey_2020,Palastro:2020gcl,Jolly:2020,simpson2022spatiotemporal,Ramsey_2022,pigeon2024ultrabroadband,Formanek:2023zia,ramsey2023exact,gong2024laser}. Experimental configurations and conceptual proposals for producing FF pulses have used chromatic optics \cite{Froula_2018,Turnbull_2018,Jolly:2020}, axiparabola-echelon pairs \cite{Palastro:2020gcl,smartsev2019axiparabola,oubrerie2022axiparabola,pigeon2024ultrabroadband}, and nonlinear optical processes \cite{simpson2020nonlinear,simpson2022spatiotemporal}. In the context of high-field physics, a FF pulse with an intensity peak that moves at the speed of light in the opposite direction of its phase fronts allows for extended interaction times with high-energy particles that collide head-on with the phase fronts. This configuration was first suggested to accentuate signatures of the transverse-formation-length in the emission of radiation \cite{DiPiazza:2020wxp}. The same configuration was later proposed in Refs. \cite{Formanek:2021bpw,Formanek:2023mkx} to facilitate the detection of classical radiation-reaction and vacuum birefringence. The observed effect in both cases, i.e., the classical energy loss of the electron beam and polarization rotation of an x-ray probe beam, scaled with the energy of the laser pulse \cite{DiPiazza:2011tq,Burton:2014wsa,Gonoskov:2021hwf,Fedotov:2022ely}. Thus, the extended interaction time afforded by the FF enabled the use of much lower and more controllable powers than ultrashort, stationary-focus Gaussian (SFG) pulses. The magnitude of the effect produced by the FF and SFG was, however, the same. 

In this letter, we show that a FF pulse can significantly enhance observable effects of quantum radiation reaction when compared to a SFG with the same energy. More specifically, FF pulses result in a greater loss of electron energy and a larger yield of photons in the 1--20 MeV range than SFG pulses. This is because the electron energy loss scales more favorably with the interaction time than the laser intensity in the quantum regime ($\chi \gtrsim 1$), while the photon yield scales more favorably with the interaction time in both the classical and quantum regimes. A larger yield of photons in the 1--20 MeV range could impact several applications, including radiosurgery \cite{girolami1996photon, weeks1997compton}, photo-transmutation for treatment of long-lived nuclear waste and production of medical isotopes \cite{wang2017transmutation,irani2014gamma,luo2016production}, and investigations into the structure of materials using nuclear resonance fluorescence \cite{kneissl1996investigation,albert2011design}. While many methods have been proposed to increase the yield of high-energy photons (see e.g., \cite{powers2014quasi,vyskovcil2020inverse,luedtke2021creating,morris2021highly,hadjisolomou2023gamma,yu2024compact}), here, we show something qualitatively different: how space-time structured light can significantly enhance the emission probability of the underlying fundamental process. 

\section{Radiation energy loss}
The basic SFQED process describing radiation by an ultrarelativistic electron beam in the field of a laser pulse is nonlinear Compton scattering. Here, this process will be investigated within the locally-constant field approximation (LCFA) \cite{DiPiazza:2011tq,Burton:2014wsa,Gonoskov:2021hwf,Fedotov:2022ely}. The LCFA is a useful approximation to study SFQED phenomena because it allows one to compute the probability of a process in an arbitrary electromagnetic field using a known expression for the probability in a constant crossed field. A constant crossed field is a space and time-independent electromagnetic field $(\bm{E},\bm{B})$ that satisfies $|\bm{E}|=|\bm{B}|$ and $\bm{E}\cdot\bm{B}=0$ \cite{Ritus:1985vta,Reiss:1962nhe,Baier:1998vh}. The LCFA is valid provided that: (a) ultrarelativistic charges move fast enough to experience an arbitrary electromagnetic field as an approximate crossed field in their rest frame \cite{Jackson_b_1975} and (b) the formation length of a process like nonlinear Compton scattering is much smaller than the wavelength of the laser pulse $\lambda_0=2\pi/\omega_0$, where $\omega_0$ is the central angular frequency of the pulse. This latter condition is typically satisfied when the dimensionless field strength $\xi_0 = |e|E_0/m\omega_0 \gg 1$, where $E_0$ is the electric field amplitude of the pulse \cite{Ritus:1985vta,Baier:1998vh}.

Within the LCFA, the average energy $\mathcal{E}_\gamma$ radiated by an electron per unit time in a background electromagnetic field through emission of a single photon can be calculated using the Baier-Katkov interpolation formula \cite{Baier:1998vh}
\begin{equation}\label{eq:interpol}
		\frac{d\mathcal{E}_\gamma}{dt} \approx 
		\frac{2\alpha m^2\chi^2/3}{[1 + 4.8(1 + \chi)\ln(1 + 1.7\chi) + 2.44\chi^2]^{2/3}},
\end{equation}
where $\alpha \approx 1/137$ is the fine-structure constant and $\chi$ is evaluated using the local values of $p_\nu$ and $F^{\mu\nu}$. The interpolation has an accuracy better than 2\% for all values of $\chi$ \cite{Baier:1998vh}. In the regime under consideration where the electron is ultrarelativistic ($\xi_0\gg 1$ and $\chi\sim 1$), the likelihood of multiple photon emission within a formation length is suppressed. Further, the work done by a laser pulse within a formation length is negligible compared to the energy lost to photon emission. Thus, the quantity $-d\mathcal{E}_\gamma/dt$ is approximately equal to the average energy lost by an electron per unit time, i.e., $d\mathcal{E}/dt=-d\mathcal{E}_\gamma/dt$, where $\mathcal{E}=\mathcal{E}(t)$ is the electron energy at time $t$ \cite{Glauber_1951,Di_Piazza_2010}. Equation (\ref{eq:interpol}) can then be used to calculate the relative electron energy loss $\zeta$ defined as
\begin{equation}\label{eq:E_loss}
	\zeta \equiv \frac{\mathcal{E}_0 - \mathcal{E}_F}{\mathcal{E}_0},
\end{equation}
where $\mathcal{E}_0$ is the initial energy of an electron and $\mathcal{E}_F$ its energy after interacting with the laser pulse. Equation \ref{eq:interpol} applies to a single electron but can be interpreted as the continuous, average energy loss per-electron for an ensemble of electrons sharing the same initial condition. This is verified in Section I of the Supplemental Material (SM) \cite{supplemental} by comparing the results of Eq. \eqref{eq:interpol} to a stochastic quantum mechanical approach (see also Ref. \cite{niel2018quantum}). 

\begin{figure}
	\begin{center}
		\includegraphics[width=\linewidth]{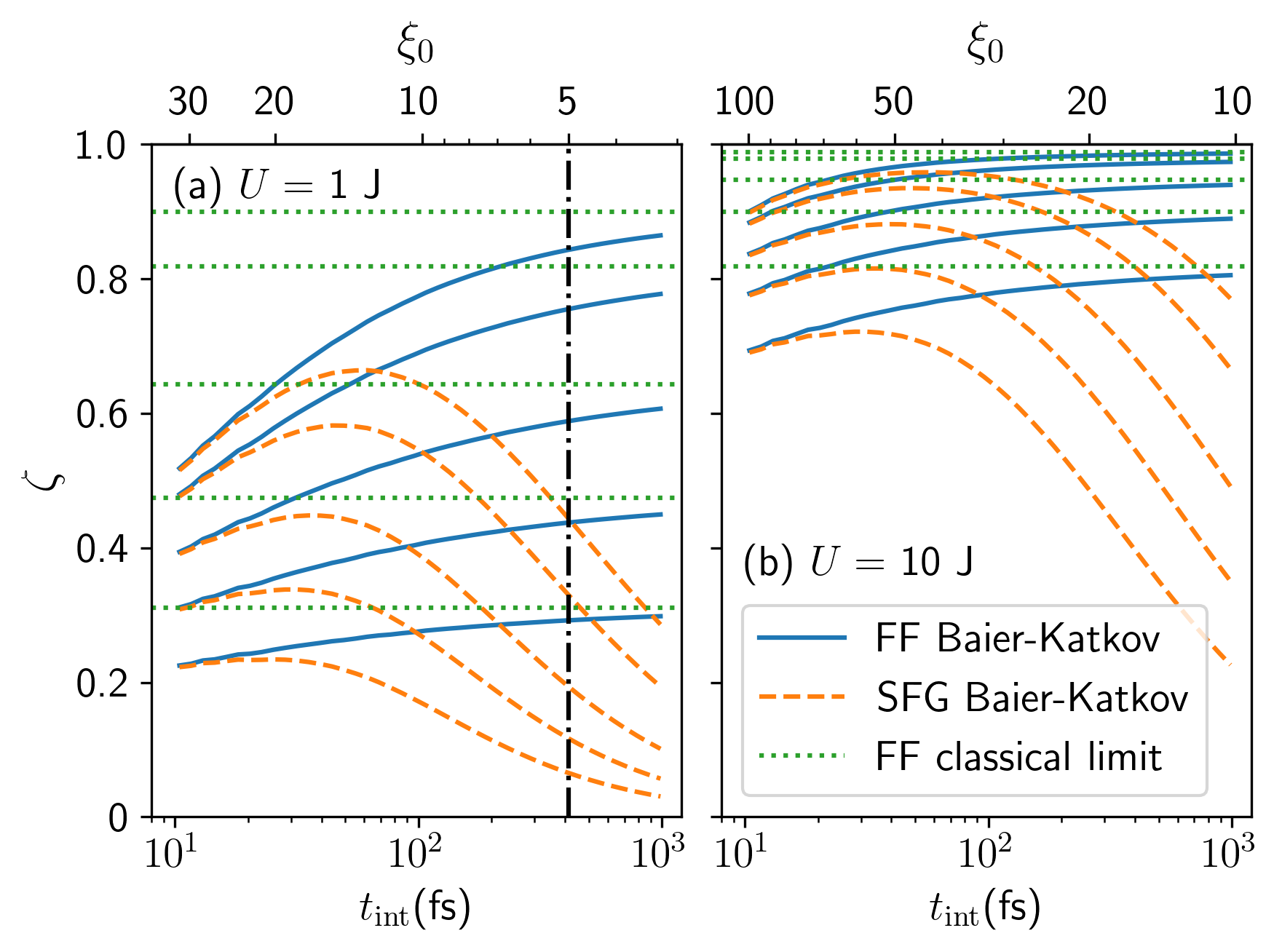}
	\end{center}
        \vspace{-2em}
	\caption{\label{fig:loss_gamma} Relative energy loss $\zeta$ as a function of the interaction time for an electron colliding head-on and on-axis with equal-energy FF and SFG pulses. From top to bottom, the lines correspond to initial electron energies of 10, 5, 2, 1, and 0.5 GeV . Both the FF and SFG pulses had a $\lambda_0 = 1\ \mu$m wavelength, $\sigma_0 = 1.5\ \mu$m focal spot size, maximum quantum nonlinearity parameter $\chi_0 \approx 0.1 \xi_0 \mathcal{E}_0$(10 GeV), and energies of either (a) $U=1\;\text{J}$ or (b) $U=10\;\text{J}$.
    For the SFG,  $z_R = 42$ fs. To the right of the dash-dotted vertical line, the results may have errors larger than 4\% due to the breakdown of the LCFA.}
\end{figure}
The advantage of using a FF pulse in place of a SFG pulse with the same energy will be demonstrated by numerically integrating Eq. (\ref{eq:interpol}). An analytical treatment is also possible by making a few simplifying assumptions: (1) $\mathcal{E}\gg m\xi_0$, which guarantees that, for an electron initially counterpropagating with respect to the laser pulse, the transverse electron motion and pulse structure can be ignored; (2) The focal point of the laser pulse is located at $x=y=0$ and the electron moves along the negative $z$ axis at an ultrarelativistic velocity such that $z(t)\approx z(0) -t$; (3) The temporal profile of the pulse has a square shape equal to unity with a duration $2t_\text{int}$, which intersects the electron from $z=t_{\text{int}}/2$ to $z=-t_{\text{int}}/2$. 

In the classical regime ($\chi \ll 1$), Eq. (\ref{eq:interpol}) can be approximated as $d\mathcal{E}/dt \approx -2\alpha m^2 \chi^2/3$. When interacting with the FF pulse, the electron experiences a field with a near-constant amplitude $\xi_0$ such that $\chi_\text{FF}(t) \approx 2 \mathcal{E}(t) \omega_0 \xi_0 |\sin(2\omega_0 t)|/ m^2$. The relative energy loss \cite{Formanek:2021bpw} is then 
\begin{equation}\label{eq:classical}
	\zeta_\text{C,FF} = \frac{\kappa_\text{C,FF}}{1 + \kappa_\text{C,FF}}, \:\:
	\kappa_\text{C,FF} \approx 2\frac{U(\text{J})\mathcal{E}_0(\text{GeV})}{[\sigma_0(\mu\text{m})]^2}\,,
\end{equation}
where $U = \pi E_0^2 \sigma_0^2 t_\text{int}/2$ is the pulse energy \cite{Esarey:1993zz} and $\sigma_0$ its focal spot size. This result shows that the energy loss in the classical regime depends on the fluence of the laser pulse, i.e., it is $\propto U/\sigma_0^2$, but is independent of the interaction time $t_\text{int}$ \footnote{Note that subleading oscillatory terms were ignored.}. For a SFG pulse, the finite Rayleigh range $z_R = \omega_0 \sigma_0^2/2$ can be taken into account by substituting $\xi_0 \rightarrow \xi_0/\sqrt{1+z^2/z_R^2}=\xi_0/\sqrt{1+t^2/z_R^2}$ in the expression for $\chi$, such that $\chi_\text{SFG}(t) \approx 2 \mathcal{E}(t) \omega_0 \xi_0|\sin(2\omega_0 t)|/m^2\sqrt{1+t^2/z_R^2}$. As a result,  $\kappa_\text{C,SFG}=\kappa_\text{C,FF}\arctan(\rho)/\rho$, with $\rho \equiv t_\text{int}/2z_R$ \cite{Formanek:2021bpw}. Thus, for a fixed laser energy, a FF pulse and a short SFG pulse with $\rho\ll 1$ induce the same energy loss in the classical regime.

In the quantum regime ($\chi\gtrsim 1$), the scalings change because the emission rate [Eq. (\ref{eq:interpol})] has a weaker dependence on the field amplitude than in the classical regime. This suggests that, in the quantum regime, a long FF pulse can induce more energy loss  than a short SFG pulse with the same energy. This reasoning can be verified analytically for FF pulses in the limit $\chi \gg 1$, where $d\mathcal{E}/dt \propto -\chi^{2/3}$ and
\begin{equation}
\mathcal{E}_0^{1/3} - \mathcal{E}_F^{1/3} \propto U^{1/3} t_\text{int}^{2/3}.
\end{equation}
Thus, for a fixed pulse energy $U$, higher energy losses can be achieved by extending the interaction time.

\begin{figure}
	\begin{center}
		\includegraphics[width=\linewidth]{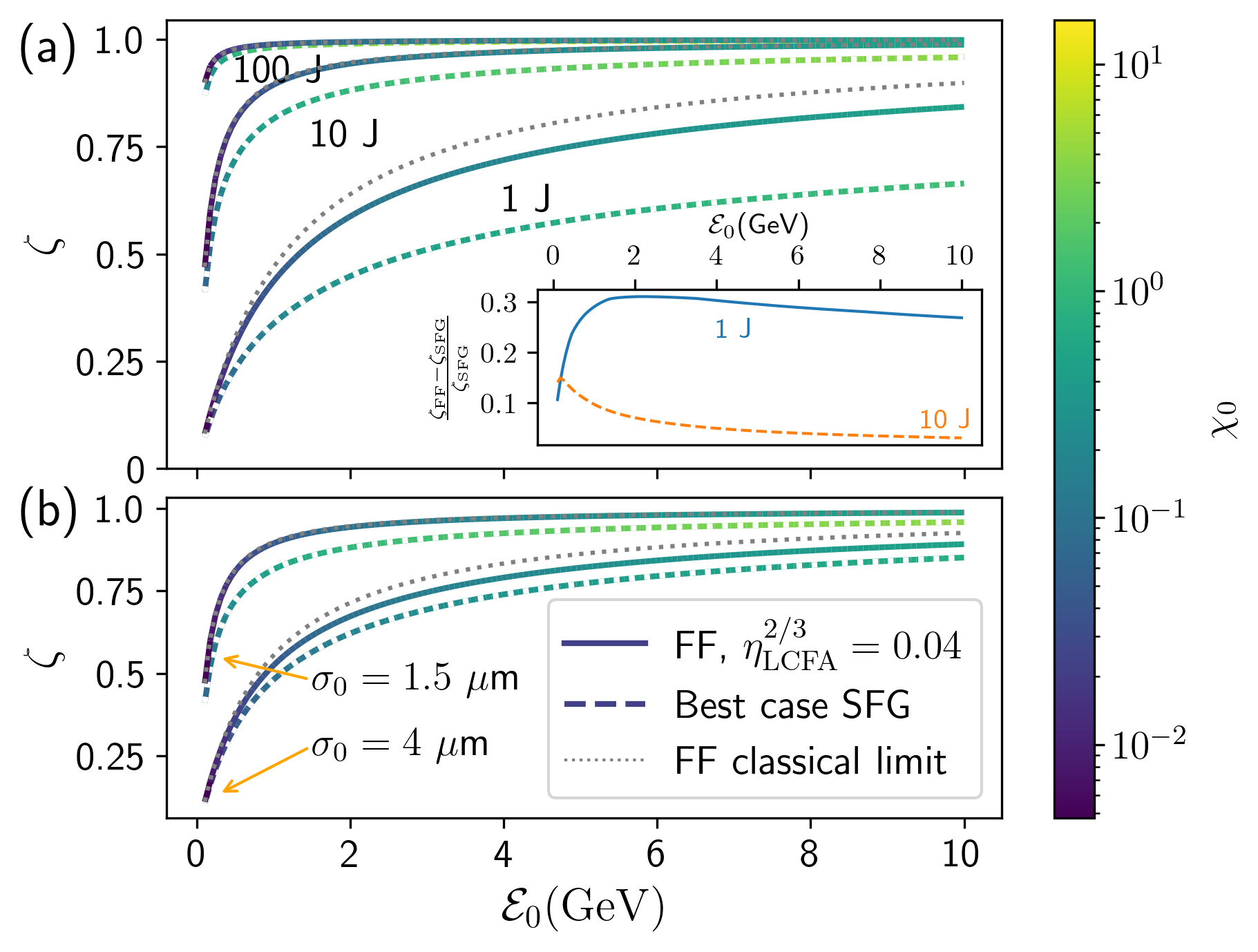}
	\end{center}
        \vspace{-2em}
	\caption{\label{fig:best_case_gamma}Relative energy loss $\zeta$ as a function of the initial electron energy for electrons colliding head-on and on-axis with equal-energy FF and SFG pulses. In (a), the pulses have an energy $U=$ 1 J, 10 J, or 100 J and a spot size $\sigma_0=$1.5 $\mu$m. In (b), $U=$ 10 J and the spot size is $\sigma_0=$ 1.5 $\mu$m or 4 $\mu$m. The color scale shows the maximum quantum nonlinearity parameter. Above $\mathcal{E}\approx$ 3 GeV, both the FF and SFG are in the quantum regime. The inset shows the relative improvement in the energy loss afforded by the FF pulses. At $U=$ 1 J, the FF results in ${\sim}30\%$ more loss than the SFG. At $U=$ 10 J, both the FF and SFG begin to approach $100\%$ loss, which limits the relative improvement achievable with a FF pulse.
    }
\end{figure}

Figure \ref{fig:loss_gamma} confirms the above expectations by showing the relative energy loss as a function of the interaction time for electrons colliding head-on with either a FF pulse (blue solid lines) or a SFG pulse (dashed orange lines) with energies of (a) 1 J or (b) 10 J. The solid and dashed lines were obtained by numerically integrating Eq.~(\ref{eq:interpol}), with $\chi$ equal to $\chi_\text{FF}$ or $\chi_\text{SFG}$, respectively. In both cases, the maximum value of the quantum nonlinearity parameter is $\chi_0 = 2 \mathcal{E}_0 \omega_0 \xi_0 /m^2\approx 0.1\,\xi_0\mathcal{E}_0(10\;\text{GeV})$. For $U=$ 1 J, the maximum energy loss achieved by a FF pulse is approximately $30\%$ higher than that of an SFG pulse regardless of the initial electron energy. For $U=$ 10 J and $\mathcal{E}_0 =$ 0.5 GeV, the maximum energy loss achieved by a FF pulse is $12\%$ higher. In this case and for larger electron energies, the enhanced energy loss in the FF pulses is reduced because the electron loses almost all of its energy whether it interacts with a FF or SFG pulse (see also the inset of Fig. \ref{fig:best_case_gamma}a discussed below).

Figure \ref{fig:loss_gamma} also shows that the energy loss in SFG pulses is suppressed for either very short pulses ($\rho \ll 1$) or very long pulses ($\rho \gg 1$). Shorter duration SFG pulses have higher peak intensities, but, as discussed above, the emission rate has a weaker scaling with intensity than it does with duration in the quantum regime \cite{niel2018quantum}. With long duration SFG pulses, the electron spends a large fraction of the interaction time outside of the confocal region where the intensity is lower. This contrasts with FF pulses where the electron spends the entire interaction near the focus where the intensity is high. As a result, the energy loss increases monotonically with the interaction time until it approaches the classical limit [green dotted lines evaluated using Eq. (\ref{eq:classical})]: for fixed pulse energy, $\chi_0 \propto t_\text{int}^{-1/2}$. The energy loss in the FF and SFG pulses coincide at short interaction times because the two pulses have the same amplitude and the finite Rayleigh range of the SFG pulse does not play a role ($\rho\ll 1$). 

The results presented in Fig. \ref{fig:loss_gamma} were verified by evolving electrons in the full FF and SFG fields with a Monte Carlo radiation emission calculation that takes into account the stochastic nature of multiple photon emission (see Section I of SM \cite{supplemental}). Notwithstanding, some remarks are in order to address the validity of the LCFA. The LCFA overestimates the amplitude of the exact QED emission spectrum for sufficiently low photon energies \cite{Di_Piazza_2018,Di_Piazza_2019,Ilderton_2019_b}. The key parameter is \cite{Di_Piazza_2018}
\begin{equation}
\label{eta}
    \eta_\text{LCFA} = \frac{\mathcal{E} -\omega}{\omega} \frac{\chi}{\xi^3}\,,
\end{equation}
where $\omega$ is the energy of the emitted photon and $\xi$ and $\chi$ are the classical and quantum nonlinearity parameters at the instant of emission. The condition $\eta_\text{LCFA}^{2/3}\ll 1$ ensures that corrections to the LCFA are small \cite{Di_Piazza_2019}. Equation \eqref{eq:interpol} is derived by integrating the emitted photon spectrum over the photon energy. For $\chi \ll 1$, the largest contribution to the integral comes from energies $\omega\lesssim \chi \mathcal{E}\ll\mathcal{E}$; for $\chi \sim 1$, the largest contribution comes from energies $\omega\lesssim \mathcal{E}$. As a result, the validity of the LCFA is determined by the low-energy photons and the condition $\eta_\text{LCFA}^{2/3} \sim 1/\xi^2 \ll 1$. The vertical line in Fig. \ref{fig:loss_gamma} corresponds to the choice $\xi_0 = 5$ \cite{Blackburn:2018sfn} so that $\eta_\text{LCFA}^{2/3}= 1/\xi_0^2=0.04$. Note that the expression for the classical energy loss does not rely on the validity of the LCFA, which explains why in the FF case the curves tend to the classical result as the interaction time increases ($\chi_0 \propto t_\text{int}^{-1/2}$ for fixed energy). 

To further illustrate the advantage of FF pulses in the quantum regime of radiation reaction, Fig. \ref{fig:best_case_gamma} compares the maximum energy loss induced by FF pulses (with the longest $t_\mathrm{int}$ satisfying $\eta_\text{LCFA}^{2/3}\sim 1/\xi_0^2 < 0.04$) to that of SFG pulses (the maxima of the orange dashed curves in Fig. \ref{fig:loss_gamma}). For almost all initial electron energies $\mathcal{E}_0$, the FF pulses (solid lines) result in a larger loss of electron energy than the SFG pulses (dashed lines). However, as the pulse energy increases, the electron energy loss for both the FF and SFG pulses approaches unity (Fig. \ref{fig:best_case_gamma}a), which limits the relative advantage of the FF pulses (Fig. \ref{fig:best_case_gamma}a inset). In addition, both pulses achieve higher energy losses with smaller spot sizes (Fig. \ref{fig:best_case_gamma}b) [see also Eq. (\ref{eq:classical}) for the classical regime]. Thus, defocusing an SFG pulse to increase the interaction time would not improve its results. 

\begin{figure}
	\begin{center}
		\includegraphics[width=\linewidth]{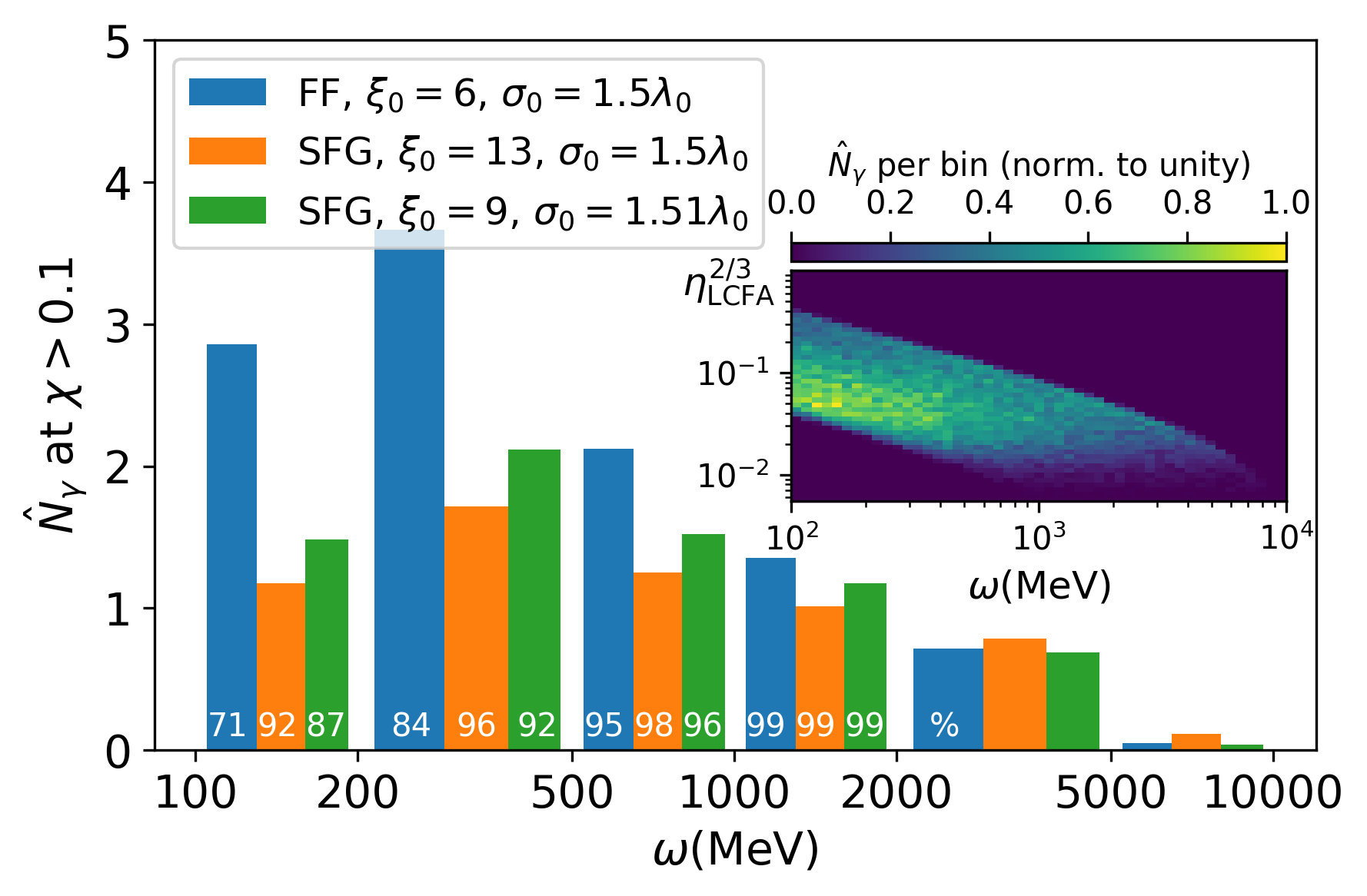}
	\end{center}
        \vspace{-2em}
	\caption{\label{fig:chi_histogram} Number of photons emitted per electron at $\chi > 0.1$ from the on-axis collision of $\mathcal{E}_0 = 10$ GeV electrons with 1~J FF and SFG laser pulses. The inset shows the number of photons emitted as a function of energy and $\eta_\text{LCFA}^{2/3}$ for the FF pulse. The numbers in white indicate what percentage of photons in the given range were emitted at $\chi > 0.1$.}
    \end{figure}

\subsection{Photon yield}
The saturation of the electron energy loss at higher pulse energies suggests that an observable without a theoretical upper limit would better elucidate the benefits of FF pulses in these conditions. One such observable, that is also important for applications, is the total number of emitted photons. Before presenting numerical calculations of the photon number, it is instructive to examine its scaling behavior to explain the advantages of a FF pulse compared to a SFG pulse. Within the LFCA, the probability $P$ of single photon emission per unit time can be calculated using the interpolation formula \cite{Baier:1998vh} 
\begin{equation}\label{eq:probability}
    \frac{dP}{dt} \approx \frac{1.45 \alpha m^2\chi}{\mathcal{E} (1 + 7.2\chi + \chi^2)^{1/6}}\,,
\end{equation}
which is better than 2\% accurate for all values of $\chi$. Multiplying the time-integrated probability by the number of electrons colliding with the laser pulse provides an estimate of the average number of emitted photons. Provided that the energy loss is accounted for as discussed below Eq. (\ref{eq:interpol}), this estimate is quantitatively accurate if higher-order processes, such as multiple photon emission within a formation length can be neglected.

In the classical regime ($\chi\ll 1$), Eq. \eqref{eq:probability} scales as $dP/dt \propto \chi/\mathcal{E}$; while in the quantum regime ($\chi\gg 1$), $dP/dt \propto \chi^{2/3}/\mathcal{E}$. These proportionalities indicate that the emission probability $P$ scales more favorably with the interaction time than the laser intensity in both regimes. Thus, for fixed collision parameters (i.e., pulse energy, electron energy, focusing), FF pulses have an advantage over SFG pulses regardless of $\chi$. 

To demonstrate this, we first consider the most favorable conditions for a SFG pulse, which is the classical regime where $\chi\ll 1$. Maximizing the emission probability for a SFG pulse requires matching the interaction time $t_\text{int}$ to the Rayleigh range $z_R$, so that electrons spend much of the interaction near the focus where the intensity is high. The optimum is $t_\text{int} \approx 5z_R - 7z_R$, which yields $P_\text{C,SFG}\propto U^{1/2}$ (see Section II of SM \cite{supplemental}). In a FF pulse, the focus moves with the electrons, so the entire interaction time is spent near the focus. As a result, the interaction time is independent of the FF Rayleigh range $z_{R,\text{FF}} \equiv \omega_0 \sigma_{0}^2$. In this case, $P_\text{C,FF} \propto (t_\text{int}/z_{R,\text{FF}})^{1/2} U^{1/2}$, such that the average number of emitted photons increases with the interaction time. The ratio $P_\text{C,\text{FF}}/P_\text{C,SFG} \propto (t_\text{int}/z_{R,\text{FF}})^{1/2}$ reveals that, for fixed laser pulse energy, a FF pulse will have an advantage over a SFG pulse when the interaction time of the FF is longer than its Rayleigh range. In the deep quantum regime ($\chi \gg 1$) the scaling is even more favorable, starting as $P_\text{Q,FF} \propto U^{1/3}t_\text{int}^{2/3}$ before asymptoting to the classical scaling as the electron loses energy. 

\paragraph{Quantum enhancement.} Figure \ref{fig:chi_histogram} demonstrates that a FF pulse can yield more photons emitted in the quantum regime than SFG pulses with the same energy. 
In the on-axis collision of 10 GeV electrons with a 1 J, $\xi_0 = 6$ FF pulse (as considered in Fig. \ref{fig:loss_gamma}a), 10.0 photons were emitted at $\chi > 0.1$ per electron in the energy range 100 MeV -- 2 GeV compared to 5.2 photons per electron with a 1 J, $\xi_0 = 13$ SFG pulse (maximum energy loss from Fig. \ref{fig:loss_gamma}a) and compared to 6.3 photons per electron in a 1 J, $\xi_0 = 9$ SFG pulse. The amplitude, spot size, and interaction time ($t_\text{int} = 128$ fs) of the $\xi_0 = 9$ SFG pulse were optimized to achieve the highest photon yield with $\chi > 0.1$ in the 100 MeV -- 2 GeV range. The electrons lost 84\%, 64\%, and 59\% of their initial energy in the FF, $\xi_0=13$ SFG, and $\xi_0=9$ SFG pulses, respectively. 
The electron energy distributions had initial standard deviation widths of 1\%, which were broadened by quantum photon emission to 39\%, 50\%, and 38\% with maximum quantum parameters $\chi_0 \approx 0.59$, $1.23$, and $0.83$ \cite{Neitz_2013,Neitz_2014,Vranic:2015sft}.
The photon spectra were generated using the Monte Carlo code described in Section I of SM \cite{supplemental}. Photons emitted with energies below 100 MeV in the FF collision are not reported because they do not meet the requirement for satisfying the LCFA approximation discussed below Eq.\eqref{eta}.

\begin{figure}
	\begin{center}
            \includegraphics[width=\linewidth]{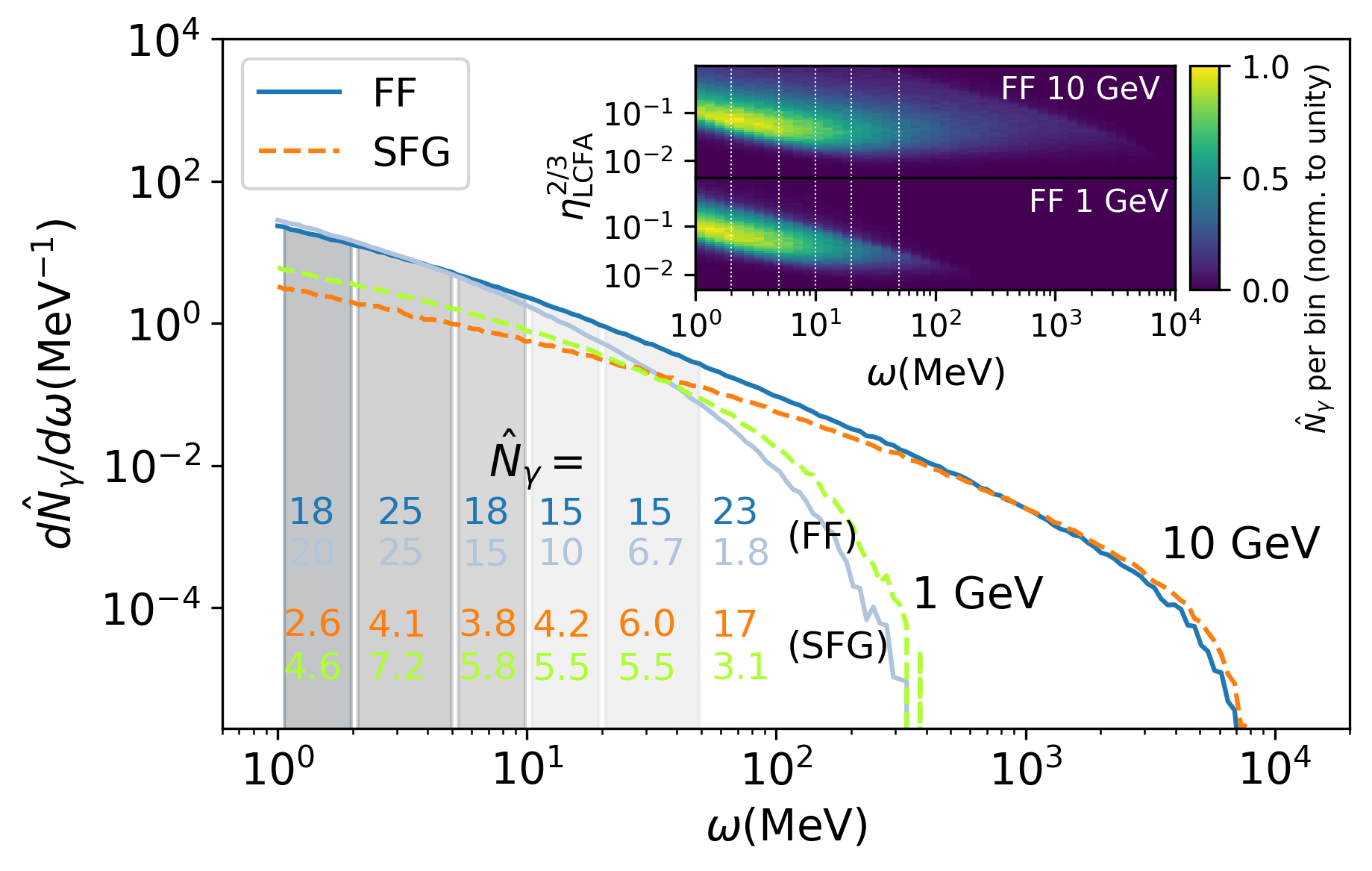}
	\end{center}
        \vspace{-2em}
	\caption{\label{fig:photon_spectrum} Spectra of emitted photons from the head-on collision of $\mathcal{E}_0 = 10$ GeV and 1 GeV electrons with $U = 10$ J laser pulses. The inset shows the number of photons emitted as a function of energy and $\eta_\text{LCFA}^{2/3}$ for the FF pulse collisions. The number of photons emitted per electron in each shaded interval is displayed in the bottom left. The FF pulse (solid lines) results in approximately five-times and three-times more photons in the 1--20 MeV range than the SFG pulse (dashed lines) for 10 and 1 GeV electrons, respectively.}
\end{figure}

\paragraph{Photon yield increase.} The enhanced photon yield in FF pulses is confirmed in Fig. \ref{fig:photon_spectrum}, which compares the number of photons $\hat{N}_{\gamma}$ emitted per unit photon energy per electron from the collision of $\mathcal{E}_0 = $ 10 GeV and 1 GeV electrons with $U = $ 10 J pulses. The parameters were chosen to be experimentally feasible \cite{los2024observation}: the wavelength and spot size of the laser pulses were $\lambda_0 = 1\ \mu$m and $\sigma_0 = 2\ \mu$m, respectively, and the electrons were initialized with normally distributed transverse positions with a variance of $0.75\ \mu$m, a divergence of 1 mrad, and a length of $5\lambda_0$. The remaining parameters were determined by maximizing the photon yields from the SFG and FF pulses while still satisfying the LCFA: The FF pulse (solid line) had $\xi_0 = 6$ and $t_\text{int} = 1.61$ ps, and the SFG pulse (dashed line) had $\xi_0 = 16$ and $t_\text{int} = 227$ fs. Note that the choice of $\xi_0 = 6$ in the FF case (as opposed to $\xi_0 = 5$ used in Fig. \ref{fig:loss_gamma}) ensures the validity of the LCFA for off-axis electrons that experience lower field strengths. 

Figure \ref{fig:photon_spectrum} shows that the FF pulse colliding with 10 GeV electrons results in more photons per electron than the SFG pulse by a factor of 7 in the 1 -- 2 MeV range, a factor of 5 in the 2 -- 10 MeV, and a factor of 4 in 10 -- 20 MeV range. Based on these results, the collision of a 100 pC beam of $\mathcal{E}_0$ = 10 GeV electrons with a FF pulse could produce $5\times 10^{10}$ photons in the 1 -- 20 MeV range, compared to only $9\times 10^9$ photons with a SFG pulse. 

Despite its much lower amplitude ($\xi_0 = 6$), the FF pulse produces a substantial number of high-energy photons ($\omega_0 > 20$ MeV) comparable to that produced by the higher amplitude ($\xi_0 = 16$) SFG pulse. However, most of the photons are in the MeV range. For 10 GeV electrons, there are two-times more photons in the 1--20 MeV range (accounting for 5\% of $\mathcal{E}_0$ on average) than above 20 MeV (accounting for 92\% of $\mathcal{E}_0$). The maximum quantum parameter in the FF collision is $\chi_0 \approx 0.6$ and many photons are emitted in the quantum regime of radiation reaction. 

For a FF pulse colliding with 1 GeV electrons, there are eight-times more photons in the 1--20 MeV range than above 20 MeV. The 1--20 MeV photons account for 36\% of $\mathcal{E}_0$ on average, while the photons above 20 MeV account for 34\%. The higher photon yield produced by the FF is due to a larger number of photons below 20 MeV. As $\chi_0 \approx 0.06$, the classical scaling applies, and the 3 times increase in the number of 1 -- 20 MeV photons is comparable to the $\sim$2.7 times overall increase predicted by the classical scaling, $dP/dt \propto \chi/\mathcal{E}$.

\section{Simulation validation} 

Because the advantage of the FF pulse occurs in the low-energy part of the spectrum, it is important to verify that the breakdown of the LCFA at low photon energies does not invalidate the results [see Eq. (\ref{eta}) and Refs. \cite{Di_Piazza_2018,Di_Piazza_2019,Ilderton_2019_b}]. To assess possible deviations from the LCFA in this part of the spectrum, simulations were performed with the particle-in-cell (PIC) code Smilei \cite{DEROUILLAT2018351} including the SFQED-tookit \cite{Montefiori:2023mkc}, which implements corrections beyond the LCFA \cite{Di_Piazza_2019}. These simulations are in excellent agreement with the Monte Carlo results presented in Figs. \ref{fig:chi_histogram} and \ref{fig:photon_spectrum} (see Section III of SM \cite{supplemental}). As additional verification, the insets of Figs. \ref{fig:chi_histogram} and \ref{fig:photon_spectrum} show that almost all photons in the studied ranges are emitted with $\eta_\text{LCFA}^{2/3} < 0.1$.

Finally, the Monte Carlo code used to generate the photon spectra in Figs. \ref{fig:chi_histogram} and \ref{fig:photon_spectrum} does not account for the decay of the emitted photons into electron-positron pairs, which would also radiate \cite{mercuri2021impact}. As demonstrated in Section IV of the SM \cite{supplemental}, pair production can be safely neglected for both the SFG and FF pulses considered in Figs. \ref{fig:chi_histogram} and \ref{fig:photon_spectrum}. At higher field strengths ($\xi_0 \gtrsim 40$), pair production in a SFG pulse would deplete the most energetic photons and slightly increase the low-energy photon yield, but the overall energy loss and photon yield would be suppressed (see Section III of the SM \cite{supplemental}).

\section{Conclusions}

In conclusion, we have shown that a flying focus (FF) pulse can significantly enhance the signatures of quantum radiation reaction when compared to a stationary focus Gaussian (SFG) pulse with the same energy. First, an electron can radiate significantly more energy in a FF pulse than in a SFG pulse. Second, a FF pulse enhances the yield of emitted photons, especially in the low-energy part of the spectrum. Moreover, the number of photons emitted in the quantum regime of the interaction (with $\chi > 0.1$) can be larger when using a FF in place of a SFG pulse.
These improvements are a direct result of the energy loss in the quantum regime and the photon emission probability in both the quantum and classical regimes scaling more favorably with the interaction time than the field intensity. Thus, any Compton scattering configuration can use a FF pulse in place of a SFG pulse to exploit this advantage. Furthermore, the use of FF pulses lowers the required power and intensity, bypasses the need for larger compression gratings, and provides a more controllable environment with simpler diagnostics of the interaction field strength \cite{Popruzhenko:2023tqx}, while simultaneously enhancing the radiation signatures. 

\textit{Acknowledgments.}
The authors thank Samuele Montefiori and Matteo Tamburini for kindly sharing a development version of their SFQED-toolkit module \cite{Montefiori:2023mkc} for SMILEI and providing support. We also thank Elias Gerstmayr for discussions regarding laboratory sources of gamma rays. This project has received funding from the European Union’s Horizon Europe research and innovation program under the Marie Sk\l{}odowska-Curie grant agreement No. 101105246-STEFF. This work was supported by the Ministry of Education, Youth and Sports of the Czech Republic through the e-INFRA CZ (ID:90254). This material is based upon work supported by the U.S. Department of Energy [National Nuclear Security Administration] University of Rochester ``National Inertial Confinement Fusion Program'' under Award Number DE-NA0004144 and U.S. Department of Energy, Office of Science, under Award Number DE-SC0021057.

This report was prepared as an account of work sponsored by an agency of the United States Government. Neither the United States Government nor any agency thereof, nor any of their employees, makes any warranty, express or implied, or assumes any legal liability or responsibility for the accuracy, completeness, or usefulness of any information, apparatus, product, or process disclosed, or represents that its use would not infringe privately owned rights. Reference herein to any specific commercial product, process, or service by trade name, trademark, manufacturer, or otherwise does not necessarily constitute or imply its endorsement, recommendation, or favoring by the United States Government or any agency thereof. The views and opinions of authors expressed herein do not necessarily state or reflect those of the United States Government or any agency thereof.

\bibliography{references}



\section*{Supplementary material}

# Supplemental Material: Enhanced quantum radiation with flying-focus laser pulses

Martin S. Formanek,[1, *] John P. Palastro,[2] Dillon Ramsey,[2] and Antonino Di Piazza[3, 2]

[1]*ELI Beamlines Facility, The Extreme Light Infrastructure ERIC, 252 41 Dolní Břežany, Czech Republic*
[2]*Laboratory for Laser Energetics, University of Rochester, Rochester, New York 14623, USA*
[3]*Department of Physics and Astronomy, University of Rochester, Rochester, New York 14627, USA*

## I. VERIFICATION OF BAIER-KATKOV INTERPOLATION FORMULA

The theoretical framework of our approach is based on the Baier-Katkov interpolation formula [Eq. (1) in the main text] [1], which gives the estimate of the average energy radiated via single photon emission by an ultrarelativistic electron in a given external electromagnetic field within the locally-constant field approximation (LCFA). Under the assumption that the correlation among successively emitted photons is negligible, the average energy lost by an electron is approximately given by the energy it radiates via single photon emission [2, 3]. As we have explained in the main text, this is the case in the classical regime, where the emissions of successive photons are uncorrelated but it is not guaranteed to hold in the quantum regime. Here, we prove the validity of this approximation by using a more elaborate model than in the main text.

We have developed a code that propagates electrons one by one in external 3D electromagnetic fields. For simulation details, see Section V. In order to model the photon emission, the Monte Carlo stochastic radiation reaction module was adapted from the Smilei particle-in-cell code [4]. This module was engaged for $\chi > 10^{-4}$ and allowed a single electron to emit multiple photons by taking into account their correlations, i.e., the fact that the electron energy at each emission undergoes a recoil affecting the emission of the successive photon(s). Each photon is emitted in the direction of the electron momentum at the time of emission. For $\chi < 10^{-4}$ a classical Landau-Lifshitz radiation-reaction force was utilized, while ignoring the field-derivative term, which typically gives rise to effects smaller than quantum effects in a plane wave [5].

Figure 1 shows an excellent agreement between the average energy loss for a simulated sample of 5000 electrons initialized on the axis with no transverse momentum (discrete points) and the energy loss obtained by integrating the Baier-Katkov formula in flying-focus (FF) pulses (solid lines) or stationary-focus Gaussian (SFG) pulses (dashed lines). Thus, the single-electron radiation intensity presented in Ref. [1] can be used to predict the average energy loss of an ensemble of electrons within the LCFA approximation also in the quantum regime, as we have done in the

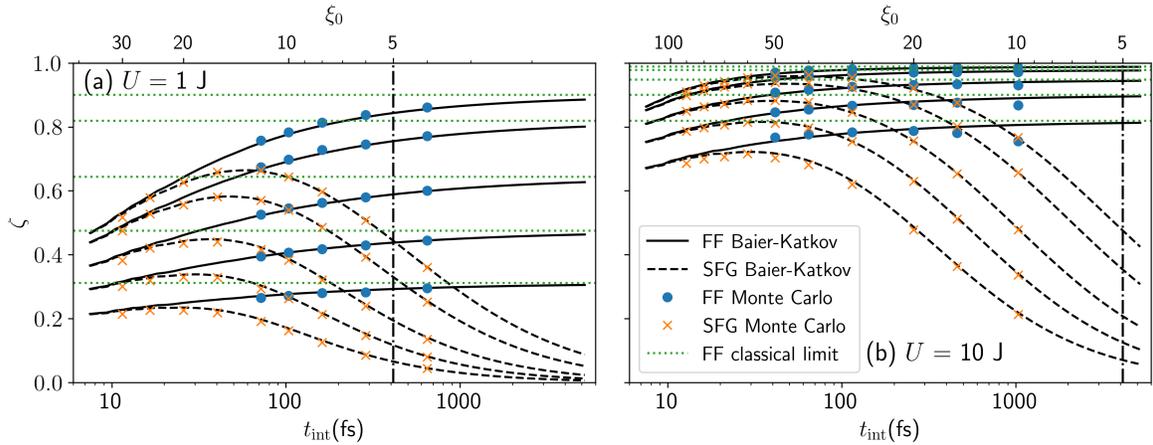

FIG. 1. Relative energy loss $\zeta$ as a function of the interaction time for electrons colliding head-on and on-axis with equal energy FF and SFG pulses of (a) $U = 1$ J and (b) $U = 10$ J. From top to bottom the lines correspond to initial electron energies of 10, 5, 2, 1 and 0.5 GeV. The solid/dash-dotted lines are results of numerical integration of the Baier-Katkov interpolation formula for radiated power. The discrete points are results of 3D particle propagation in analytically prescribed laser fields with quantum Monte Carlo radiation reaction. In the regions on the right of the dash-dotted vertical line the LCFA results may undergo corrections larger than $1/5^2 \approx 4\%$ [see the discussion below Eq. (5) in the main text].

* martin.formanek@eli-beams.eu

main text (see also [6]). Only FF pulses longer than the FF Rayleigh range were considered because for shorter pulses the FF regime would not be beneficial.

## II. INTEGRATED PROBABILITY OF THE PHOTON EMISSION

In this section the integrated probability for the single-photon emission [Eq. (6) from the main text] is evaluated in both the FF and SFG cases in the limit $\chi \ll 1$. To keep track of the variables, let us denote the FF Rayleigh range as $z_{R,FF} = \omega_0 \sigma_{0,FF}^2$, and the SFG Rayleigh range as $z_{R,SFG} = \omega_0 \sigma_{0,SFG}^2/2$ (note the factor of two difference [7]). In the FF case the integrated probability for $\chi \ll 1$ can be evaluated as

$$P_{C,FF} \approx 1.45 \alpha m^2 \int_{-t_{int}/2}^{t_{int}/2} \frac{\chi_{C,SFG}(t)}{\mathcal{E}(t)} dt = 2.9 \alpha \omega_0 \xi_0 \int_{-t_{int}/2}^{t_{int}/2} |\sin(2\omega_0 t)| dt = 2.9 \alpha \omega_0 \xi_0 t_{int} \kappa \,, \tag{1}$$

where $\kappa = t_{int}^{-1} \int_{-t_{int}/2}^{t_{int}/2} |\sin(2\omega_0 t)| dt \approx 0.64$ for $t_{int} \gg \lambda_0$. The total energy in the pulse is given by $U = \pi m^2 \omega_0^2 \xi_0^2 \sigma_0^2 t_{int}/(2e^2)$ which holds true for both FF and SFG pulses and allows us to eliminate $\xi_0$

$$P_{C,FF} \approx 2.9 \frac{\sqrt{2} e \alpha \kappa}{\sqrt{\pi} m} \frac{\sqrt{U t_{int}}}{\sigma_{0,FF}} = 2.9 \frac{\sqrt{2} e \alpha \kappa}{\sqrt{\pi} m} \sqrt{\frac{\omega_0 U t_{int}}{z_{R,FF}}} \,. \tag{2}$$

The integrated probability increases with longer interaction times or smaller spot sizes (shorter Rayleigh ranges).

In the SFG case the integral for probability in the $\chi \ll 1$ regime is

$$\begin{aligned} P_{C,SFG} &\approx 1.45 \alpha m^2 \int_{-t_{int}/2}^{t_{int}/2} \frac{\chi_{C,SFG}(t)}{\mathcal{E}(t)} dt = 2.9 \alpha \omega_0 \xi_0 \int_{-t_{int}/2}^{t_{int}/2} \frac{|\sin(2\omega_0 t)| dt}{\sqrt{1 + t^2/z_{R,SFG}^2}} \\ &< 2.9 \alpha \omega_0 \xi_0 \int_{-t_{int}/2}^{t_{int}/2} \frac{dt}{\sqrt{1 + t^2/z_{R,SFG}^2}} = 2.9 \alpha \omega_0 \xi_0 2 z_{R,SFG} \mathrm{arcsinh}\left(\frac{t_{int}}{2 z_{R,SFG}}\right) \,, \end{aligned} \tag{3}$$

where $|\sin(2\omega_0 t)|$ was replaced by unity as an upper bound and integrated analytically. Evaluating once again $\xi_0$ in terms of the total pulse energy yields

$$P_{C,SFG} < 2.9 \frac{\sqrt{2} e \alpha}{\sqrt{\pi} m} \sqrt{\omega_0 U} \sqrt{\frac{2 z_{R,SFG}}{t_{int}}} \mathrm{arcsinh}\left(\frac{t_{int}}{2 z_{R,SFG}}\right) < 2.9 \frac{\sqrt{2} e \alpha \Gamma}{\sqrt{\pi} m} \sqrt{\omega_0 U} \,. \tag{4}$$

The last expression is evaluated at the maximum for $\rho = t_{int}/2 z_{R,SFG} \approx 3.32$ when $\Gamma \equiv \mathrm{arcsinh}(\rho)/\sqrt{\rho} \approx 1.05$. Such upper bound estimate for the photon yield has no dependence on the interaction time or focusing. For a given pulse energy $U$ the ratio of probabilities is

$$\frac{P_{C,FF}}{P_{C,SFG}} > \frac{\kappa}{\Gamma} \sqrt{\frac{t_{int}}{z_{R,FF}}} \approx 0.62 \sqrt{\frac{t_{int}}{z_{R,FF}}} \,. \tag{5}$$

To conclude, when $\chi \ll 1$ the formula for the Baier-Katkov single photon emission probability predicts that FF pulses have an advantage as long as their interaction with the electrons is at least about 2.6 times longer than their Rayleigh range. In reality, even shorter time could be sufficient because Eq. (4) was derived as an upper bound.

## III. BEYOND THE LCFA

In order to test the effects of the corrections beyond the LCFA at low photon energies, two simulations were compared. In the first one, the laser collision with 10 GeV electrons was simulated in 2D with the particle-in-cell code Smilei [4], which implements a Monte Carlo radiation module. The second simulation used SFQED-toolkit [8] instead of the default radiation module. The SFQED-toolkit implements the corrections beyond the LCFA at low photon energies as described in Ref. [9]. For the details of the Smilei simulations see Section VI of this Supplemental material. Panels (a) and (b) in Fig. 2 show this effect on collision of 10-GeV electrons with 1-J pulses of varying pulse length and thus varying $\xi_0$. The resulting emission spectra without LCFA correction are in excellent agreement with those

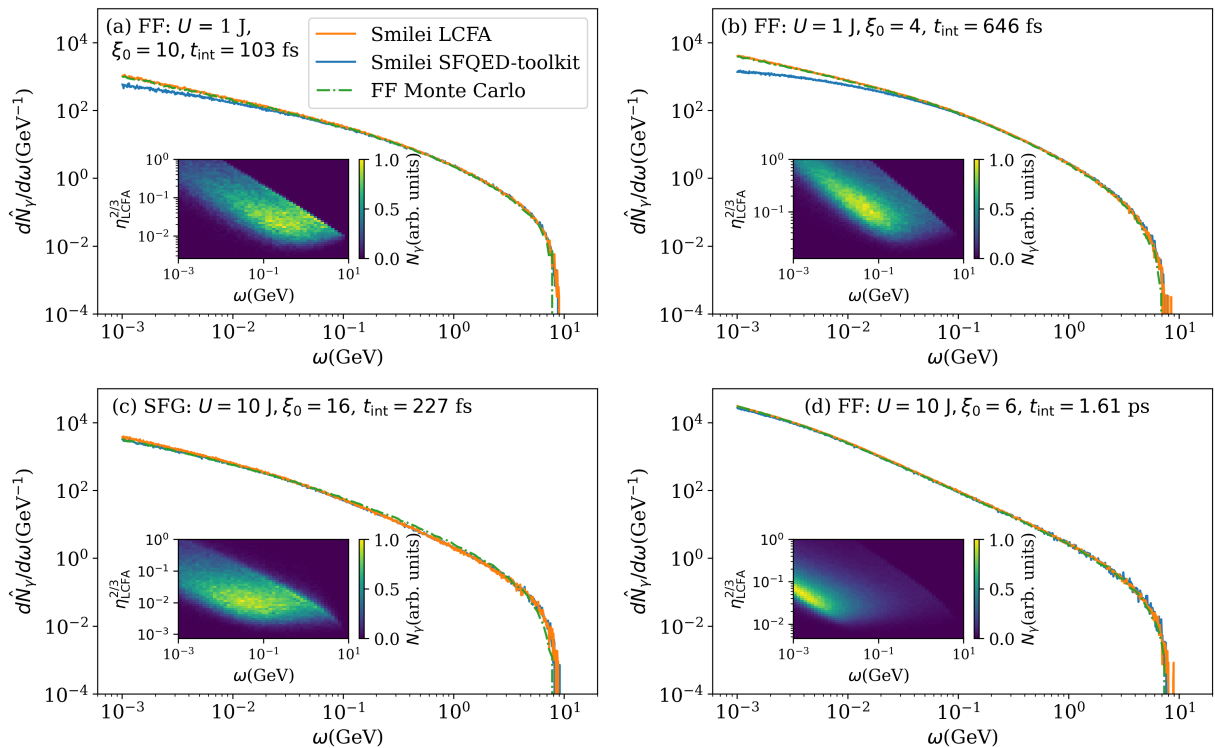

FIG. 2. Photon emission spectra for collision of 1 J [parts (a) and (b)] and 10 J [part (c)] laser pulses and 10 GeV electrons calculated with Smilei LCFA (top solid lines), Smilei beyond LCFA (bottom solid lines), and our Monte Carlo (dash-dotted lines). The insets show the parameter $\eta_{LCFA}^{2/3}$ for the emitted photons.

calculated by our Monte Carlo code with full fields for laser focusing $\sigma_0 = 1.5$ $\mu$m. The beyond-LCFA module corrects the low-energy part of the spectrum and the corrections become more prominent as the interaction time increases because $\xi_0$ then decreases. As it can be seen in the insets of Fig. 2, the corrections are important in the regions where most of the emitted photons have $\eta_{LCFa}^{2/3} > 0.1$. Nevertheless, since the corrections are sizable only for soft photons, the effect on the total energy loss is negligible. The predictions of the time integration of the Baier-Katkov formula presented in the main text, our Monte Carlo code, and Smilei with and without the beyond-LCFA corrections are all within 1% of each other.

Currently, it is not possible to simulate FF pulses in Smilei beyond the plane-wave approximation. To produce Fig. 2, FF pulses were modeled in Smilei as an ideal plane wave of duration $2t_{int}$(fs) $= U(J)/[2.15 \times 10^{-5}\sigma_0^2(\mu m)\xi_0^2]$ for $\lambda_0 = 1$ $\mu$m, which is a reasonable approximation for initially on-axis electrons, as they locally experience plane-wave-like field.

Similar comparison can be made for the spectra presented in the main text. Figure 2c displays the emission spectrum for a collision of 10-J SFG laser pulses focused to the spot size $\sigma_0 = 2$ $\mu$m with a 10-GeV electron beam with initial variance of the radial position from axis $\sigma_e = 0.75$ $\mu$m, 1 mrad divergence, and 5 $\mu$m length. In this case, beyond-LCFA corrections do not affect the photon spectrum for emitted photons with energies above 1 MeV. The FF pulse in Fig. 2d was again replaced in Smilei by an ideal plane wave and compared to Monte Carlo simulation with electrons initialized on the $z$-axis. Since most of the photons are emitted with $\eta_{LCFA}^{2/3} < 0.1$, the spectra again match extremely well. The more realistic electron beam used in the main text has electrons initialized off-axis, off-center and with non-zero divergence. Such electrons experience lower field strength which increases $\eta_{LCFA}^{2/3}$ for the emitted photons. However, as it can be seen in Fig. 3 of the main text, most of the emitted photons stay in the region $\eta_{LCFA}^{2/3} < 0.1$, where beyond-LCFA corrections are negligible.

## IV. ELECTRON-POSITRON PAIR PRODUCTION

The Monte Carlo code used in this study to obtain the photon spectra does not take into account the interaction of the emitted photons with the background laser field. However, high-energy photons produced via nonlinear Compton

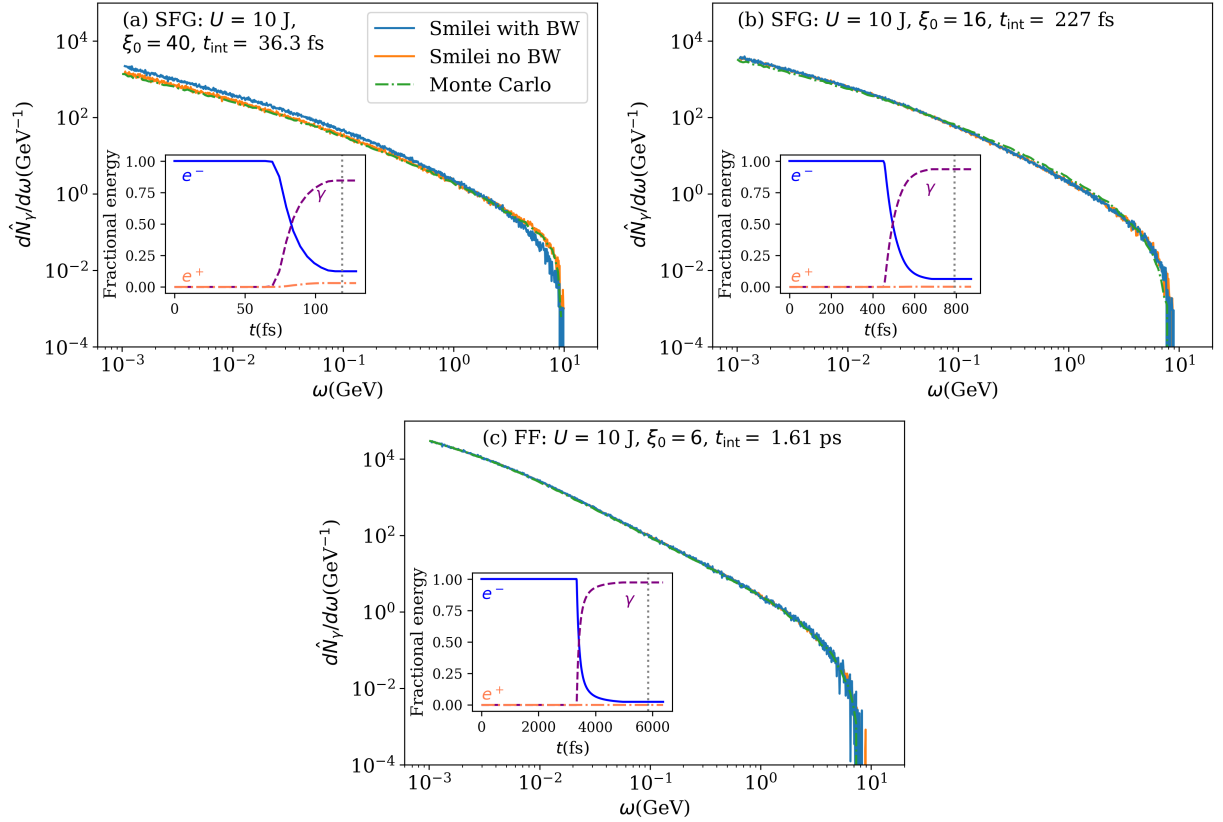

FIG. 3. Photon emission spectra without (solid line) and with (dashed line) Breit-Wheeler (BW) pair production module in Smilei compared with our Monte Carlo simulation (dash-dotted line). The insets show the time evolution of the electron (solid line), photon (dashed line), and positron (dash-dotted line) fractional energy as a function of time for the Smilei with BW simulation. The vertical dotted lines denote the times when the photon spectra are plotted.

scattering could decay into electron-positron pairs via nonlinear Breit-Wheeler (BW) pair production, and the decay products could radiate and contribute to the spectra, as well. The quantity determining the probability of the photon decay is the quantum nonlinearity parameter of the emitted photons $\chi_\gamma$. In the background plane wave field and for photons moving in the opposite direction as the laser phase velocity, it is $\chi_\gamma \approx 2\xi_0\omega_0\omega/m^2$. The emitted photon energies, especially those where the FF has an advantage, are sufficiently low that pair production can be safely neglected. For example, for photons with energies in the range 1 - 20 MeV colliding with a laser field with $\lambda_0 = 1$ $\mu$m and $\xi_0 = 6$, it is $\chi_\gamma = 5 \times 10^{-5}$ - $1 \times 10^{-3}$.

In the higher intensity SFG pulses $\chi_\gamma$ could be orders of magnitude higher and electron-positron pairs are expected to be produced. In order to test the importance of this effect, Smilei simulations of collisions between 10-GeV electrons and 10-J laser pulses with (a) $\xi_0 = 40$, $t_{int} = 36.3$ fs (b) $\xi_0 = 16$, $t_{int} = 227$ fs laser pulses were carried out (see the corresponding panels in Fig. 3). In both cases, the laser pulse was focused to a spot size of $\sigma_0 = 2$ $\mu$m and the initial radial electron spread had variance $\sigma_e = 0.75$ $\mu$m with 1 mrad divergence and $5\lambda_0$ beam length. Figure 3a shows that for the high field strength $\xi_0 = 40$ the abundance of the highest-energy photons decreases as expected, whereas the number of photons in the 1 - 20 MeV region of interest increases only slightly. In total 0.30 electron-positron pairs are produced per initial electron. In the optimal SFG case, maximizing the photon yield (presented in the main paper) pair production can be safely neglected as only 0.017 pairs per initial electron are produced (see Fig. 3b).

Finally, for the FF case, the background field was modeled in Smilei as a plane-wave field with $\xi_0 = 6$ and $t_{int} = 1.61$ ps and compared with Monte Carlo simulation with electrons initialized on-axis. Figure 3c confirms numerically that nonlinear BW pair production does not play any role as only $3.2 \times 10^{-4}$ pairs per initial electron are produced. In the main text for the Monte Carlo simulations, a more realistic electron beam was used with nonzero length, containing electrons initialized off-axis, off-center, and with nonzero divergence. Such electrons experience lower field strengths and BW pair production will contribute even less.

## V. MONTE CARLO SIMULATION PARAMETERS

Our Monte Carlo code for single-particle propagation uses a Runge-Kutta 4th order particle pusher [10] with a time step of $0.02\ \omega_0^{-1}$. For the FF field, a fully analytical prescription from SM of Ref. [7] was implemented. The leading-order paraxial approximation was used for the SFG fields. For all simulations the laser pulses had a central wavelength of $\lambda_0 = 1\ \mu$m. In order to smoothly switch on the fields, the code uses a fifth-order polynomial envelope ramp up (see SM of Ref. [7]) corresponding to either 4 wavelengths or 3% duration of the interaction time (whichever is larger). To compensate for the length of the ramp, half of it is considered as a part of the interaction time.

For the test of Baier-Katkov formula for on-axis collisions the laser spot size was set to $\sigma_0 = 1.5\ \mu$m. The main propagation direction of electromagnetic fields was in the positive $z$ direction. The initial position of all electrons was on the optical axis, which coincides with the $z$ axis, and all electrons were initialized with the same momentum strictly along the negative $z$.

For the simulations mimicking the realistic experimental conditions from Ref. [11] the laser spot size was $\sigma_0 = 2\mu$m. The electron beam was initialized with $0.75\ \mu$m variance in the initial distance from the optical axis. The beam divergence was set to 1 mrad and longitudinal beam length to 5 $\mu$m. The longitudinal distribution of the electron beam was modelled by the prescription from the SM of Ref. [12]. In the SFG case the electron beam was initially longitudinally positioned so that its center interacts with the laser pulse symmetrically around its focus. In the FF case the beam center coincides with the position of the peak intensity of the flying focus.

## VI. SMILEI SIMULATION PARAMETERS

For all Smilei simulations a 2D grid with spatial resolution of 32 points per laser wavelength was employed. The time step was chosen to be 135 steps per laser period, which is 67% of the Courant-Friedrich-Lewy maximal time step. The laser pulse of length $2t_{\text{int}}$ was injected from the left side of the simulation box ($z = 0$) and the electron beam from the right side (with initial position of the longitudinal center of the beam at $z = 4t_{\text{int}}$). That way the beam center starts interacting with the laser pulse at $z = 2t_{\text{int}}$ for the duration $t_{\text{int}}$. The "vay" pusher was employed to propagate the particles except for the beyond-LCFA calculations which use "borisby" pusher. The simulation box uses absorbing Silver-Mueller boundaries for the injecting sides of the box and periodic boundaries parallel to the optical axis.

For the SFG Smilei simulations, the laser was a 2D Gaussian focused at $z = 1.5t_{\text{int}}$ such that the interaction occurs symmetrically around the focus. The electron beam was initialized as 10,000 macro-electrons with density $n_e = 10^{-5}n_{\text{crit}}$ ($n_{\text{crit}} = 1.11 \times 10^{27}$ m$^{-3}$) with a longitudinal distribution given by the SM of Ref. [12]. The initial position of the macro-electrons was chosen to ensure that variance of their radial position at $t = 2t_{\text{int}}$ is $0.75\ \mu$m with divergence 1 mrad. The width of the simulation box was chosen to be $1.5t_{\text{int}}$ to fully resolve the focus of the laser injected from the boundary.

For the FF Smilei simulations an ideal electromagnetic plane wave was used as an approximation of the FF pulse (see above). The width of the simulation box was 8 $\mu$m. The electron beam was initialized in a thin slice (fraction of $\lambda_0$) at the right edge of the box with zero divergence.



\end{document}